\documentclass[epj, final]{svjour}
\usepackage{graphics}
\usepackage{amsfonts}
\usepackage{amsmath}
\usepackage{amssymb}
\usepackage{graphicx,color}
\usepackage{rotating}
\usepackage{hyperref}
\usepackage{dcolumn}
\usepackage{bm}
\usepackage[T1]{fontenc}
\usepackage[utf8]{inputenc}
\usepackage{subfig}
\usepackage{multirow}
\RequirePackage[numbers,sort&compress]{natbib}
\RequirePackage{flushend}
\begin{document}
\title{Quark matter revisited with non extensive MIT bag model}
\author{Pedro H. G. Cardoso\inst{1} 
	\and Tiago Nunes da Silva\inst{1} 
	\and Airton Deppman\inst{2}
	\and Débora P. Menezes\inst{1}
}                   
\institute{Departamento de Física, CFM, Universidade Federal de Santa Catarina, 88040-900, Florianópolis, Brazil
		\and
		Instituto de Fí­sica da Universidade de São Paulo, Rua do Matão, 187 - Travessa R, São Paulo, Brazil}
\offprints{Débora P. Menezes}
\mail{debora.p.m@ufsc.br}
\date{Received: date / Revised version: date}
\abstract{
In this work we revisit the MIT bag model to describe quark matter within both the usual Fermi-Dirac and the Tsallis statistics. We verify the effects of the non-additivity of the latter
by analysing two different pictures: the first order phase transition of the QCD phase diagram and stellar matter properties. While, the QCD phase diagram is visually affected by the Tsallis statistics, the resulting effects on quark star macroscopic properties are barely noticed.
} 
\maketitle
\section{Introduction}
\label{intro}
Although quantum chromodynamics (QCD) is {\it the} theory underlying the physics described by the strong interaction, its solution is far from being possible. This fact led to the developments of two complementary theoretical approaches: lattice quantum chromodynamics (LQCD) and effective models, both of them aiming to describe the QCD phase diagram. At this moment, due to computational limitations and numerical difficulties, as the sign problem, for instance, LQCD can only cover a small fraction of the QCD phase diagram, restricted to zero and very low chemical potentials \cite{Meyer:2015wax}. On the other hand, effective models rely largely on the mean field approximation and not well known hyperonic potentials, depending on which part of the QCD phase diagram one wants to study. Recent investigations taking into account a large number of relativistic models \cite{Dutra:2014qga} have shown how fragile and model dependent the results can be \cite{Dutra:2015hxa, PhysRevC.95.025201}.

In view of these obstacles, alternative phenomenological approaches increase in importance. Among those phenomenological attempts to describe high energy data, the power-law associated to Tsallis statistics is widely used \cite{Cabibbo:1975ig,Frautschi:1971ij,Deppman:2016fxs, Marques:2015mwa, Wilk:2013jsa, Zborovsky:2006nh, Wong:2013sca, Wilk:2008ue, Azmi:2015xqa, Azmi:2014dwa}. This approach considers that the highly excited system formed in high energy collisions follows Tsallis statistics instead of Boltzmann statistics, with the entropic index $q$ as a measure of the deviation respect to the latter theory \cite{Tsallis:1987eu}.

Although the first applications of Tsallis statistics in high energy physics were mainly empirical \cite{Beck:2000nz},  it was shown later that a generalisation of the self-consistent thermodynamics \cite{Deppman:2012us} exists, as proposed by Hagedorn \cite{Hagedorn:1965st}, but taking into account the non extensive behaviour of the Tsallis statistics. This non extensive statistics predicts a limiting (or critical) temperature as well as a parameter $q_o$ characteristic for all hadrons. Several analyses of transverse momentum distribution from high energy collisions confirmed such predictions \cite{Bediaga:1999hv, Deppman:2012us, Deppman:2012qt, Marques:2012px, Sena:2012ds, Sena:2012hc, Marques:2015mwa}, and also an analysis of hadron mass spectrum has shown that the non extensive theory can describe very well the known hadronic states \cite{Marques:2012px}.

An important region of the QCD phase diagram is the one where a liquid-gas (type) phase transition is possible and the effects of admitting non-extensivity have already been exploited in \cite{Lavagno:2013qca}. Other important region is the one at low temperatures and high densities, which characterises compact objects, (neutron stars, pulsars, quark stars) one of the aims of the present work.

More recently, it has been shown that a fractal structure in the thermodynamic functions leads the system to be naturally described by Tsallis statistics \cite{Deppman:2016fxs}. Since the Hagedorn and Frautisch \cite{Hagedorn:1965st,Frautschi:1971ij} descriptions of hadrons already present such fractal structure, one may conclude that the use of Tsallis statistics for hadronic systems is a consequence of the well-known self-similar characteristics of the hadron structure. Such self-similar feature of hadrons is found in experimental data \cite{Wilk:2013jsa,Zborovsky:2006nh} through different and complementary aspects. The present work presents an attempt to include such self-similarity in the MIT bag model for hadron structure. 

We next concentrate our attention on quark matter described by the simplest possible relativistic model, the MIT bag model \cite{Chodos:1974je}, which was developed at the Massachusetts Institute of Technology in the 70s. We first check how the QCD phase diagram is modified once the Fermi statistics is replaced by the Tsallis statistics and then apply the model to the description of quark stars, which are compact stellar objects constituted of deconfined quark matter. 

In previous works \cite{Lavagno:2011zm, Menezes:2014wqa}, neutron star macroscopic properties have already been investigated with a relativistic model that is popular for this purpose, known as Boguta-Bodmer \cite{Boguta:1977xi} or non-linear Walecka model \cite{Serot:1984ey}. Within this context, non-extensive statistics can be interpreted as an alternative to account for the not always well justified assumptions underlying mean field approximations \cite{Rozynek:2015zca}. 
Usually, non-extensivity (or rather {\it non-additivity}, as we prefer to put it) is incorporated by simply replacing the usual Boltzmann-Gibbs \cite{Megias:2015fra} or Fermi-Dirac statistics  \cite{Lavagno:2011zm, Menezes:2014wqa,Rozynek:2015zca} without affecting the original structure of the models under investigation. A general conclusion was that a $q$ value larger than one produces families of stars with slightly larger maximum masses. No conclusive statement could be made as far as radii were concerned. In \cite{Lavagno:2011zm}, $q$ values smaller than one were also used and the results went to the opposite direction, \textit{i.e.}, maximum masses were lower. Once the entropy was fixed at one of the commonly used snapshots of the stellar evolution, it was seen that the star internal temperature tended to decrease with the increase of the $q$ parameter. Walecka type models contain two distinct parts in their Lagrangian density: a Fermi kinetic contribution and a part that mimics the strong interaction by coupling mesons to hadrons. It is fair to say that it is not easy to see how non-additivity affects each contribution separately. In \cite{Menezes:2014wqa}, an attempt was made by analysing a free-Fermi gas at a fixed temperature (very academic and unrealistic picture) modified to incorporate Tsallis statistics with $q$ values larger than one and the resulting maximum mass showed a behaviour opposite to the case when the Walecka model was used and based on a fixed entropy picture, an indication that the effects of the non-additivity might be larger in the interaction terms that mimics the strong interaction than in the kinetic part of the EOS. 

The model we have chosen to work with in the present study is simple enough to be compared with a free Fermi gas model because all the information on the strong interaction comes from a constant, as will be clearly shown in the next section. Hence, by investigating quark matter with the MIT bag model, our intention is to shed some light to the understanding of the effects of the Tsallis statistics on the QCD phase diagram and also on quark stars. A preliminary calculation of the effects of non-\-extensivity on relativistic Fermi gas can be seen in \cite{Megias:2015fra,Rozynek:2015cna}.

Strange (quark) stars were first proposed as a realisation of deconfined quark matter at very high densities, which according to the Bodmer-Witten conjecture, could be the ground state of matter
\cite{Bodmer:1971we}. This state is only attained if strange quarks are present and this assumption led to the name strange star. Since then, the hypothesis of strange matter has been tested with different quark models, mainly the MIT bag model and the Nambu-Jona-Lasinio model \cite{PeresMenezes:2005bc}, the last one also investigated within the Tsallis statistics \cite{Rozynek:2015zca}, but in a different perspective.

We organise our work as follows: section II is devoted to a short review of the the Tsallis statistics and the MIT bag model and section III to the application of the MIT bag model within the non-additivity assumption to the description of the QCD phase diagram and to quark stars. In the last section the final remarks are presented.

\section{The Formalism}
We devote this section to review the formalism underlying the use of the Tsallis statistics and to present the main equations used to describe the MIT bag model. The derivations are not included because they can be obtained from other papers in the literature, whose citations are always given.

\subsection{Tsallis statistics}
In the following a self-similar bag model is developed by using Tsallis instead of Boltzmann statistics. In this way, the self-similar structure is introduced in the MIT bag model, according to the results presented in \cite{Deppman:2016fxs}. 

The Tsallis statistics \cite{Tsallis:1987eu} was proposed by Constantino Tsallis in 1988 as a generalisation of the Boltzmann-Gibbs Statistics and its associated entropy is given by
\begin{equation}
S_{q} = k\dfrac{1 - \sum_{i=1}^W p_{i}^q}{q-1}.
\end{equation}

The Tsallis entropy $(S_{q})$ is not thought as being an universal function that is given once and for all, but it is a delicate and powerful concept to be carefully constructed for classes of systems characterised by the real parameter ``$q$''. This entropy satisfies all properties of Boltzmann-Gibbs entropy $(S_{BG})$ , except the additivity, so that
\begin{equation}
S_{q}(A+B) = S_{q}(A) + S_{q}(B) + (1-q)S_{q}(A)S_{q}(B),
\end{equation}
in such a way that when $q \rightarrow 1$, $S_{BG}$ is recovered.

According to thermodynamics, entropy needs to be extensive. This means that, for a large number $N$ of elements (probabilistically independent or not) the entropy of the system is (asymptotically) proportional to $N$. Otherwise, the entropy is non-extensive. For a system whose elements are either independent or weakly correlated, the additive entropy $S_{BG}$ is extensive, whereas the non-additive entropy $S_{q}$ $(q \neq 1)$ is non-extensive. In contrast, however, for a system whose elements are generically strongly correlated, the
additive entropy $S_{BG}$  can be non-extensive, whereas the nonaddictive entropy $S_{q}$ $(q \neq 1)$ can be extensive for special values of $q$.

We next define the q-logarithm function for particles ($+$) and antiparticles ($-$) as in \cite{Megias:2015fra}:
\begin{equation}
\begin{cases}
& \log^{(+)}_q(x)=\frac{x^{q-1}-1}{q-1}, \quad  x \geq 0,  \\
&  \log^{(-)}_q(x)=\frac{x^{1-q}-1}{1-q},  \quad x<0\,
\end{cases}
\end{equation}
is the inverse function of the $q$-exponential given by
\begin{equation}
\begin{cases}
 & e_q^{(+)}(x)=[1+(q-1)x]^{1/(q-1)} \qquad\;\;\,\,\,\,\,\,\,\,\,\,\,\,\,\,\,\,\,\,\,\,\,\,\,\,\,, \; x\geq 0\, \\
 & e_q^{(-)}(x)=\frac{1}{ e_q^{(+)}(|x|)}=[1+(1-q)x]^{1/(1-q)}\qquad\,, \; x<0\,. \label{qexp}
\end{cases}
\end{equation}

In this way
\begin{equation}
S_{q} = -k\sum_{i=1}^W p_{i}\log_{q}p_{i}.
\end{equation}

In order to introduce non-extensivity in the EOS of the MIT bag model, the starting point is the partition function \cite{Megias:2015fra} for a non extensive ideal quantum gas
\begin{equation}
\begin{split}
\log \Xi(V,T,\mu) &= -\xi V\int\dfrac{d^{3}p}{(2\pi)^{3}}\sum\limits_{r = \pm} \\
&\times\Theta(rx)\log_{q}^{-r}\left(\dfrac{e_{q}^{r}(x) - \xi}{e_{q}^{r}(x)}\right),
\end{split}
\end{equation}
where $x= \beta(E - \mu)$, $\xi = \pm 1$ for bosons and fermions respectively, and $\Theta $ is the step function.

The distribution and entropy density functions for fer\-mions can be obtained through the relations
\begin{equation}
\left<N\right> = \beta^{-1}\dfrac{\partial}{\partial\mu}\log\Xi_{q}\bigg|_{\beta},
\end{equation}
\begin{equation}
S = -\beta^{2}\dfrac{\partial}{\partial\mu}\left(\dfrac{\log\Xi_{q}}{\beta}\right)\bigg|_{\beta},
\end{equation}
so that, the distribution functions are
\begin{equation}
\begin{cases}
& n_q^{(+)}(x)=\frac{1}{(e_q^{(+)}(x) +1)^q}, \quad \,\,\,\,\, x \geq 0,\\
& n_q^{(-)}(x)=\frac{1}{(e_q^{(-)}(x) +1)^{2-q}},\quad x<0 ,
\end{cases}
\label{eq:n}
\end{equation}
and the  entropy density
\begin{equation}
\begin{split}
 {\cal S} &= \frac{1}{\pi^2(q-1)} \sum_j \sum_{r=\pm} \int p^2 dp  \, \Theta(r x_j) r \\
&\qquad \times \bigg[ 1- n_q^{(r)}(x_j) - \left(1 - ({n}_q^{(r)}(x_j))^{1/\tilde{q}}\right)^{\tilde{q}} \bigg]\,,  \label{entropy}
\end{split}
\end{equation}
where
\begin{equation}
\tilde{q}=
\begin{cases}
& q \qquad\quad\;\;\, \,,\,\,\,x\geq 0\,, \\
& 2-q \qquad \,,\,\,\,x<0\,.
\end{cases}
\end{equation}

It is important to notice that Eq.~(\ref{eq:n}) is consistently obtained from the partition function and by the optimisation of the entropy, as proposed in \cite{Megias:2015fra} under the appropriate constraints. An interesting analysis of the several non-extensive versions of a quantum gas has been recently done in ref \cite{Rozynek:2015cna}. Notice that the distribution function $n_{q}^{(-)}$ is a direct result of the application of the usual formalism of thermodynamics, and the exponent $2-q$ results from the usual calculations.

In this context, the pressure becomes
\begin{equation}
P = \frac{T}{\pi^2}  \sum_j \sum_{r=\pm} \int p^2 dp \, \Theta(r x_j) \log^{(-r)}_q\bigg( \frac{1}{1-{n}_q^{(r)}(x_j))^{1/\tilde{q}}} \bigg) \,, \label{press-tsallis}
\end{equation}
the baryonic density reads
\begin{equation}
{\cal N}=
\begin{cases}
& \frac{1}{\pi^2} \sum_j		
\int p^2 dp  \, n_q^{(+)}(x_j)\,, \quad x_j \geq 0, \\
& \frac{1}{\pi^2} \sum_j
\int p^2 dp \, n_q^{(-)}(x_j) + 2 C_n\,, \quad x_j < 0
\end{cases}
\label{dens-tsallis}
\end{equation}
with
$$
C_n= \frac{\mu_j T \sqrt{\mu_j^2 - {m_j}^2}}{2 \pi^2} \frac{(2^{q-1}+2^{1-q}-2)}{q-1} \theta( \mu_j - m_j) 
$$
and the energy density is given by:
\begin{equation}
{\cal E}=
\begin{cases}
& \frac{1}{\pi^2} \sum_j
\int p^2 dp ~E_j~~ n_q^{(+)}(x_j)\,, \quad x_j \geq 0, \\
& \frac{1}{\pi^2} \sum_j
\int p^2 dp  ~E_j~~ n_q^{(-)}(x_j) + 2 C_e\,, \quad x_j < 0,
\end{cases}
\label{ener-tsallis}
\end{equation}
with
$$
C_e = \mu_j C_n
$$
and where $x_j=\beta(E_j - \mu_j)$.

\subsection{MIT Bag Model}
In this subsection we present the equations of state (EOS) used in this work within two different statistical approaches. We describe quark matter with the help of the MIT bag model \cite{Chodos:1974je}, which mimics the confinement of quarks in a volume space delimited by a certain pressure.

It is interesting, in the present context, to recall that one of the first conjectures about a phase transition between confined and deconfined regimes of hadronic matter was proposed by Cabibbo and Parisi \cite{Cabibbo:1975ig} as a way to interpret the Hagedorn limiting temperature. Inside the bag, a constant positive potential energy per unit of volume, the so-called \textit{Bag constant} and denoted by $B$ is necessary so that the bag can be created and kept in the vacuum. The energy associated with the mere presence of quarks in a volume $V$  is therefore $BV$. Inside of this volume, the moving quarks have a kinetic energy and no colour currents survive in the surface. Hence, the quarks in the interior of the bag are taken as a Fermi gas whose energy at the border of the bag is negligible when compared with the energies inside it. The Lagrangian density reads
\begin{equation}
\begin{split}
\mathcal{L} =& \left[\dfrac{i}{2}(\overline{\psi}\gamma^{\mu}\partial_{\mu}\psi-\partial_\mu\overline{\psi}\gamma^{\mu}\psi) - m\overline{\psi}\psi-B\right]\Theta(R-r) \\
&- \dfrac{1}{2}\overline{\psi}\Delta_{s}\psi,
\end{split}
\end{equation}
with
\begin{equation}
\Theta(R-r) 
\begin{cases}
& 1, \, \, \, \, r<R \\
& 0,  \, \, \, \, r>R
\end{cases}
\end{equation}
where $R$ is the radius of the bag and $\Delta_{s} = \delta(R-r)$ is a Dirac type function. Thus, inside of the bag the quarks are free (asymptotic freedom), and outside of the bag  they can no longer exist (confinement).  $\psi(x)$ is the fermionic field, and $m$ is the quark mass. The role of the  $B$ constant is to incorporate the QCD effects at large distances \cite{Bhaduri:1988gc, DeTar:1983rw}.

Assuming a simple relativistic mean filed approximation (RMF), the EOS can then be easily obtained at finite temperature and reads
\begin{align}
p =& -B + \notag \\
      &\dfrac{1}{3}\sum\limits_{i}\dfrac{\gamma_{i}}{2\pi^{2}}\int_0^\infty p^{3}\dfrac{\partial{E_{i}(p)}}{\partial{p}}[n(p,\mu_{i}) + n(p,-\mu_{i})]dp,
\label{originalp}\\
 \epsilon =& B + \sum\limits_{i}\dfrac{\gamma_{i}}{2\pi^{2}}\int_0^\infty p^{2}E_{i}(p)[n(p,\mu_{i}) + n(p,-\mu_{i})]dp,\label{originale}\\
n_{B} =& \dfrac{1}{3}\sum\limits_{i}\dfrac{\gamma_{i}}{2\pi^{2}}\int_0^\infty p^{2}[n(p,\mu_{i}) - n(p,-\mu_{i})]dp, \label{orignald}\\
s =& \dfrac{S}{V} = \left(\dfrac{\partial p}{\partial T}\right)_{V,\mu_{i}},
\label{MIT_EOS}
\end{align}
where 
\begin{equation}
E_{i}(k) = \left(m_{i}^{2} + k^{2}\right)^{1/2},\quad i=u,d,s
\end{equation}
$\gamma_{i}$ refers to the degeneracy of the system and accounts for the number of colours (3) and the spin (2), $m_{i}$ is the quark mass, $\mu_{i}$ the chemical potential and $n(k,\pm \mu_{i})$ is the statistical distribution for quarks and antiquarks given by:
\begin{align}
n (p,\mu_i)&= \dfrac{1}{\left\{1+e^{\beta(E_{i} - \mu_{i})}\right\}}, \\ 
  \bar{n (p, -\mu_i)} &= \dfrac{1}{\left\{1+e^{\beta(E_{i} + \mu_{i})}\right\}.}
\end{align}

When the MIT bag model EOS is calculated with the non-extensive statistical mechanics instead of the usual Fermi-Dirac, the EOS can be rewritten as:
\begin{equation}
\begin{split}
p  &= -B  \\
&+ \begin{cases}
\frac{T}{q-1} \sum_i \frac{\gamma_i}{2 \pi^2} \int p^2 dp \bigg[ -
\left( \frac{e_q^{+}(x_i)}{1+e_q^{+}(x_i)} \right)^{q-1} +1 \bigg] , \\ 
T \sum_i \frac{\gamma_i}{2 \pi^2} \int p^2 dp
\log^{(+)}_q(1+ e_q^{-}(x_i)^{-1}),
\label{tsallisp}
\end{cases}
\end{split}
\end{equation}
and
\begin{equation}
\epsilon =\sum_i \frac{\gamma_i}{2} {\cal E}, \,\,\,\, n_B =\dfrac{1}{3} \sum_i \frac{\gamma_i}{2} {\cal N}.  \quad \,  
\end{equation}
The factors of two incorporated in the denominators account for the compensation of the spin degeneracy, already taken into account in eq.(\ref{press-tsallis}), but reinserted in the equations above to make them uniform with the notation used in eqs.(\ref{originalp} - \ref{MIT_EOS}).

\section{Applications}
\begin{figure*}
\centering
\subfloat[]{\includegraphics[width=0.7\linewidth]{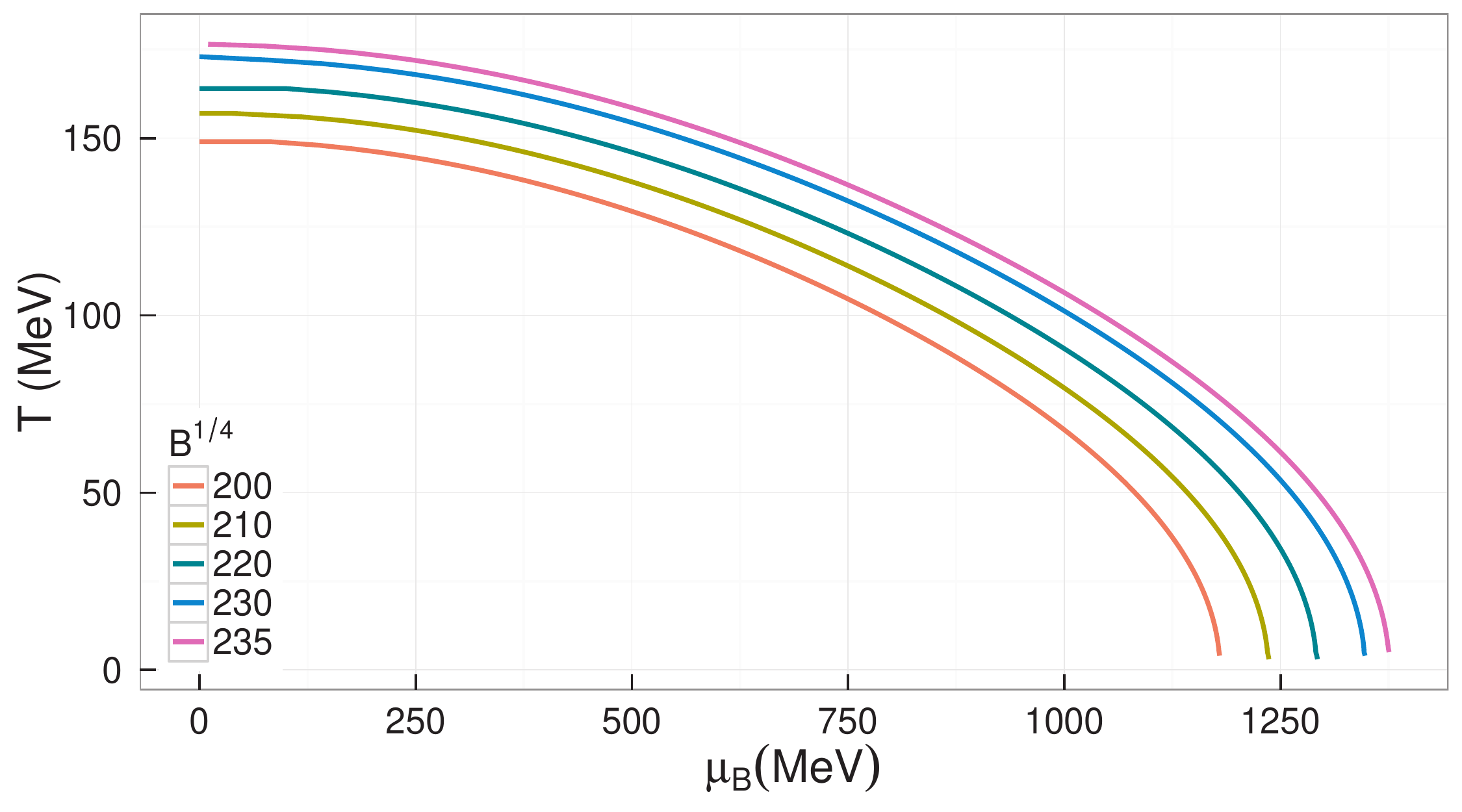}\label{fig1:a}}
\hfill
\subfloat[]{\includegraphics[width=0.7\linewidth]{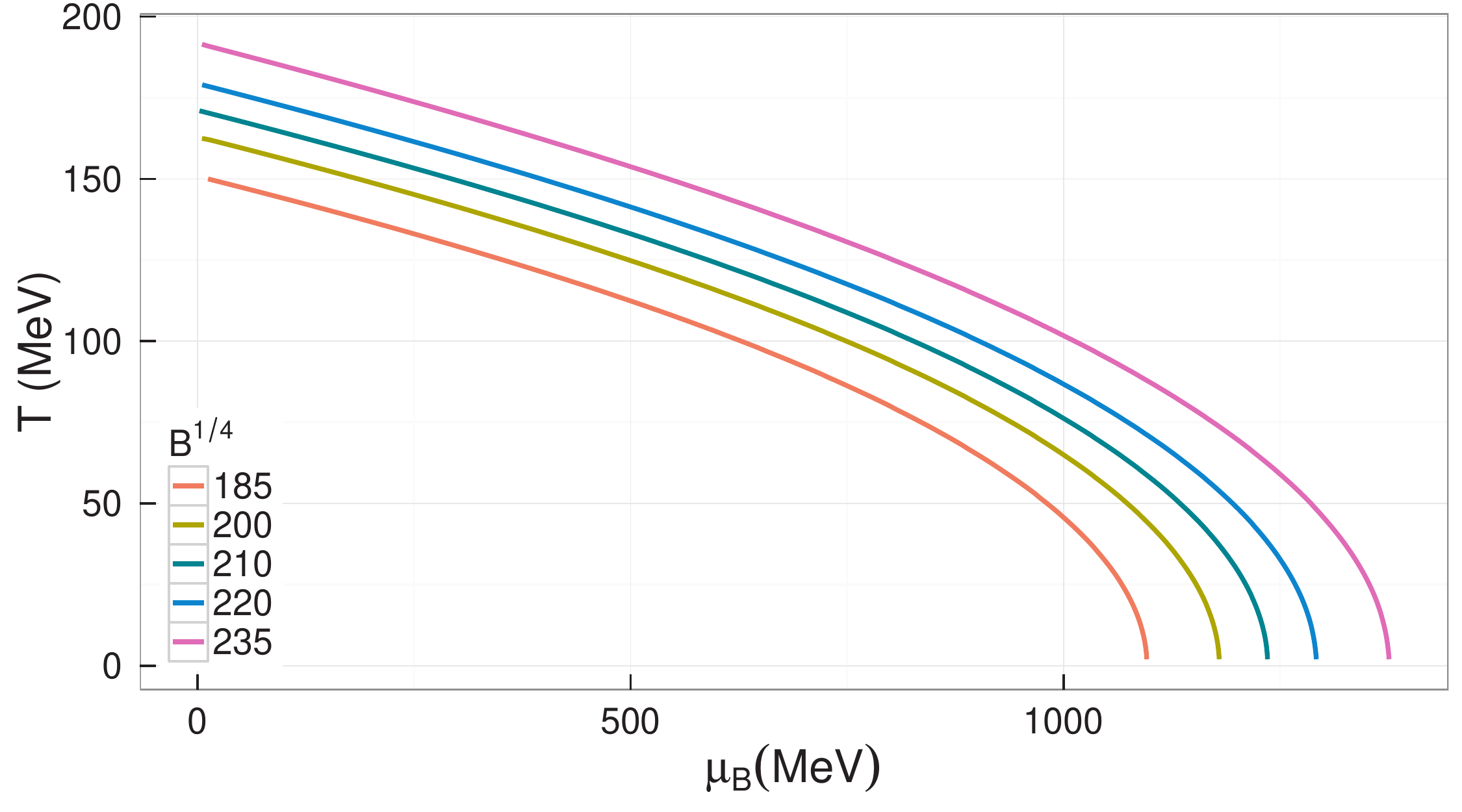}\label{fig1:b}}
\hfill
\subfloat[]{\includegraphics[width=0.7\linewidth]{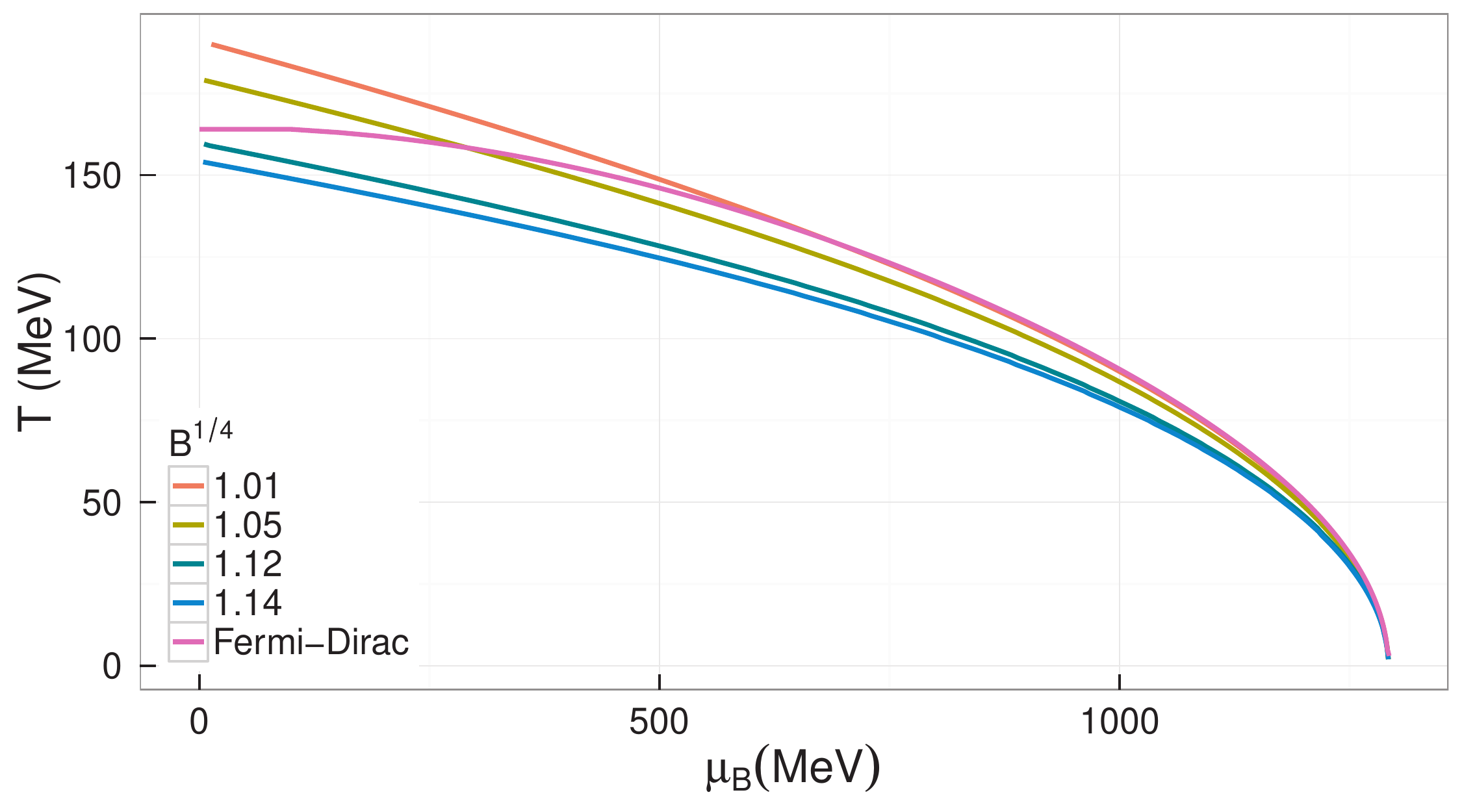}\label{fig1:c}}
\caption{QCD phase diagram using the MIT bag model for a) usual Fermi-Dirac statistics, b) Tsallis statistics with $q=1.05$ and c) Fermi-Dirac and Tsallis statistics with $B^{1/4} = 220$ MeV. } 
\label{fig1}
\end{figure*}
We next revisit two situations in which the MIT model has been used in the literature, but now also with the help of the Tsallis statistics. The first one is a simple calculation of the hadronic and quark matter separation boundary in the QCD phase diagram. The second is the calculation of compact objects macroscopic quantities, where {\it neutron stars} are assumed to be quark stars, as proposed in \cite{Bodmer:1971we}. In both cases, only $q$-values larger than 1 are considered because, from previous works, we already know that lower values make the EOS softer and we are searching for harder EOS.

\subsection{QCD diagram}
A complete analytical treatment of QCD is still beyond our reach. The only independent intrinsic scale in this theory is the dynamically generated confinement scale $\Lambda_{QCD} \sim 1$ fm$^{-1}$. As stated in the Introduction, the most reliable theoretical approach to compute thermodynamic properties of matter at finite temperatures is LQCD \cite{Meyer:2015wax}. The other possibility is the use of effective models \cite{Ferreira:2013tba,Costa:2013zca}. In both cases, the existence of a transition from a hadronic phase to a quark-gluon plasma phase at a temperature between $150$ and $200$ MeV is well established. It is worth pointing out that the use of more sophisticated models, as done in \cite{Ferreira:2013tba} may produce different (pseudo)temperatures for the chiral and deconfinement phase transitions (Table I, first line), but the values remain under the range mentioned above. Another important aspect refers to the existence and location of the critical end point (CEP), as discussed in \cite{Rozynek:2015zca,Costa:2013zca}, but this problem is out of the scope of the present paper because the model we are using here can only account for a first order phase transition all the way from high (low) to low (high) chemical potentials (temperatures).

If we use the MIT bag model to describe nuclear matter, the increase in density and temperature can make the surfaces of the bags overlap, merging them in a large area where quarks and gluons are allowed to move freely  throughout the new volume in a \textit{quark-gluon plasma} (QGP). A very naive way to obtain the separation limit of the two phases is simply by forcing eq. (\ref{originalp}) for Fermi-Dirac statistics and eq.(\ref{tsallisp}) for the Tsallis statistics to be identically zero. The quarks masses are taken as 5 MeV for the up and down quarks and 150 MeV for the strange quarks. 

Common values found in the literature for $B^{1/4}$ lie in the range $145 - 235$ MeV \cite{Rischke:1987xe}. From Fig. \ref{fig1} we see that for $B^{1/4} = 200 - 235$ MeV with the original model and for $B^{1/4} = 185 - 235$ MeV and $q=1.05$ with the Tsallis statistics, the temperature of QGP transition ($T_{c}$) lies within the expected range. This findings mean that a $q$ value larger than one plays the same role as a slightly larger $B$ value.

However, one can also see from Figs\ref{fig1} a) and \ref{fig1} b) that, for fixed values of $B$, the {\it shape} of the curves are quite different, depending on the statistics used. Moreover, if we vary the value of the parameter $q$,  the critical temperature decreases with the increase of the $q$ value. Another important aspect refers to the behaviour of the curves at low chemical potentials. When $q$ approaches 1, the curves tend to the Fermi-Dirac results, except for chemical potentials smaller than 250 MeV, when the temperatures obtained with non-extensive statistics are slightly larger than the original MIT results. It is interesting to see that the results obtained with the MIT bag model are more sensitive to non-extensivity than the ones obtained with the NJL model, as can be seen in Fig.9 of \cite{Rozynek:2015zca}. We advocate that intrinsic correlations between quarks and gluons not described by the simple MIT within the MFA can be taken into account by the $q$ value.

 In a previous work \cite{Megias:2015fra}, an attempt to describe the QCD transition line with the Tsallis statistics had already been made, by assuming two different conditions, namely, a fixed energy per particle of 1 GeV \cite{Cleymans:1998fq, Cleymans:1999st} and a fixed entropy divided by the cubic temperature equal to 5 \cite{Tawfik:2005qn}. In both cases, with $q=1.14$ the resulting transition lines were much lower than the expected chemical freeze-out line (see Fig.2 left in
\cite{Megias:2015fra}). Here, a somewhat simpler treatment brings the transition line to its correct position and the larger $q$ value ($q=1.14$) is the one that produces (pseudo) transition temperatures closer to 155 MeV, the value of the transition temperature most accepted nowadays \cite{Meyer:2015wax}. However, in the present study, the energy never reaches 1 GeV and it is not a fixed quantity, i.e., it increases with the increase of the chemical potential and decreases with the increase of the temperature. The energy per particle vary between 624 MeV  and 640 MeV for the Fermi-Dirac statistics with small variations for the non-extensive statistics, when $B^{1/4}$ is 220 MeV. As the energy per particle is given by ${\cal E}/n_B$ and ${\cal E}$ depends on the $B$ value, the final value also depends on $B$. It is interesting to see that the values we found with the naive MIT bag model for the energy density and corresponding highest temperature present in the freeze-out line go in line with the values proposed in \cite{Randrup:2009ch}.

\subsection{Stellar Matter}
\begin{figure*}
\centering
\subfloat[]{\includegraphics[width=0.7\linewidth]{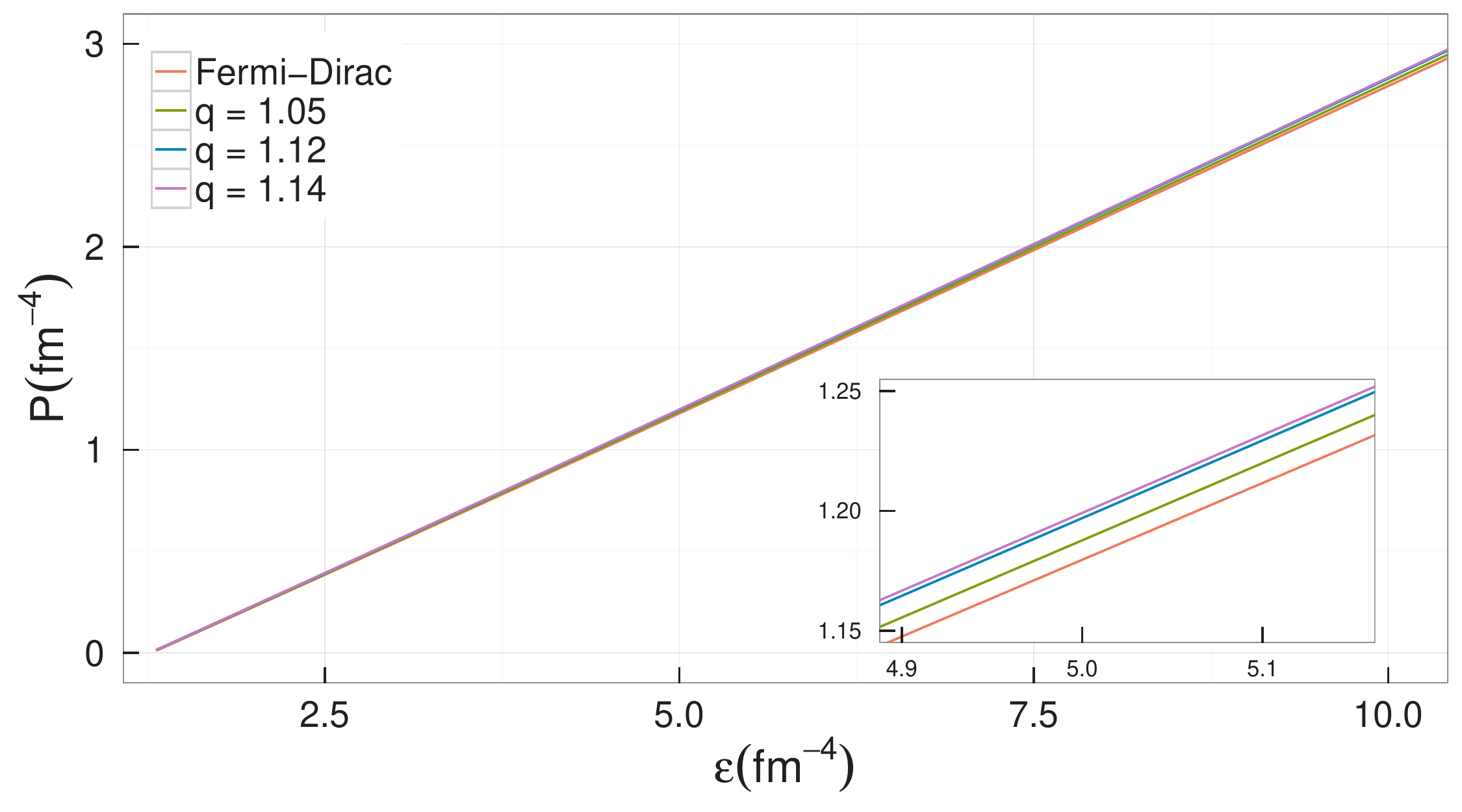}\label{fig2:a}}
\hfill
\subfloat[]{\includegraphics[width=0.7\linewidth]{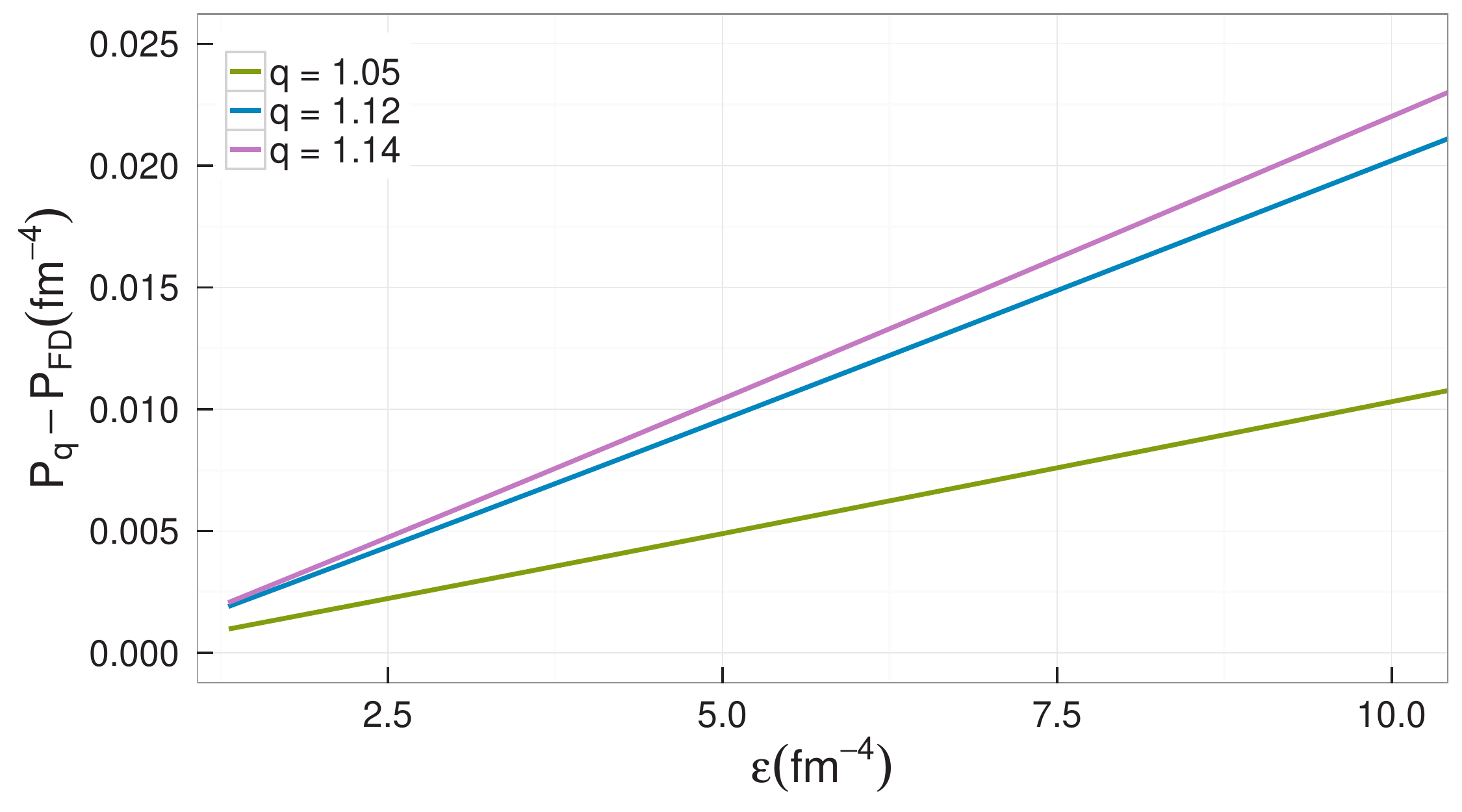}\label{fig2:b}}
\caption{a) Stellar quark matter equation of state constituted by the
  three flavours of quarks and electrons and ${\cal S}/n_B=2$, $\mu_{\nu}=0$.
b) Differences from the usual Fermi-Dirac pressure obtained with
different $q$ values versus energy density for the same snapshot.}
\label{fig2}
\end{figure*}
As pointed out in the Introduction, previous investigations of the effects of the Tsallis statistics in the description of compact objects have already been performed in \cite{Lavagno:2011zm, Menezes:2014wqa}. In both cases, neutron stars described by hadronic matter were investigated and in \cite{Menezes:2014wqa} a rapid analyses about its effect on a free gas was also done. We next study quark stars and compare our results with the ones already existing in the literature.

In stellar matter there are two conditions that have to be fulfilled, namely, charge neutrality and $\beta$-stability and they read:
\begin{equation}
\sum_j q_j n_j + \sum_l q_l n_l =0,
\end{equation}
where $q_{type}, type=j,l$ stands for the electric charge of quarks and leptons respectively and
\begin{equation}
\mu_j=q_j \mu_n - q_e (\mu_e - \mu_{\nu}), \qquad \mu_\mu=\mu_e.\label{beta}
\end{equation}
We next use the non extensive statistics also for the leptons, which enter the calculation as free particles obeying the above mentioned conditions. It is important to stress that the spin degeneracy of neutrinos, whenever they are used is 1, in contrast with the degeneracy of the other leptons, which is 2. The number of colours (3) is obviously not part of the degeneracy factor for leptons.

The entropy per particle (baryon) can be calculated through the thermodynamical expression
\begin{equation}
\frac{\cal S}{n_B}=\frac{{\cal E}+P- \sum_j \mu_j n_j}{T n_B}.
\end{equation}

One common way of studying a compact object history is by looking at three snapshots of the time evolution of a quark star in its first minutes of life, which  are given by:
\begin{itemize}
\item snapshot 1 - ${\cal S}/n_B=1$, $Y_l=0.3$,
\item snapshot 2 - ${\cal S}/n_B=2$, $\mu_{\nu}=0$,
\item snapshot 3 - ${\cal S}/n_B=0$, $\mu_{\nu}=0$,
\end{itemize}
where
\begin{equation}
Y_l=\frac{\sum_l n_l}{n_B},
\end{equation}
which, according to simulations, can reach $Y_l\simeq 0.3-0.4$. Let's point out that in snapshot 1, muons are not considered because they are either absent or very few and in snapshot 2, where neutrinos have already left the star (deleptonization era), muons are present.

In the present work we are interested in finite temperature systems and hence, all results refer to the first two snapshots. From this point on, $B^{1/4}$ is always taken equal to 145 because this value is known to be inside the stability window that satisfies the Bodmer-Witten conjecture \cite{Torres:2012xv}. In Fig.\ref{fig2}a we present the EOS obtained for the second snapshot and it is clearly very difficult to distinguish the curves (see the inset). Hence, in Fig.\ref{fig2}b  we plot the differences from the usual pressure (Fermi-Dirac) obtained with different $q$ values. At very high densities, the larger differences are smaller than 0.025\%. We do not display the curves for the first snapshot because the values are different, but the general behaviour is the same.

\begin{figure*}
\centering
\subfloat[]{\includegraphics[width=0.7\linewidth]{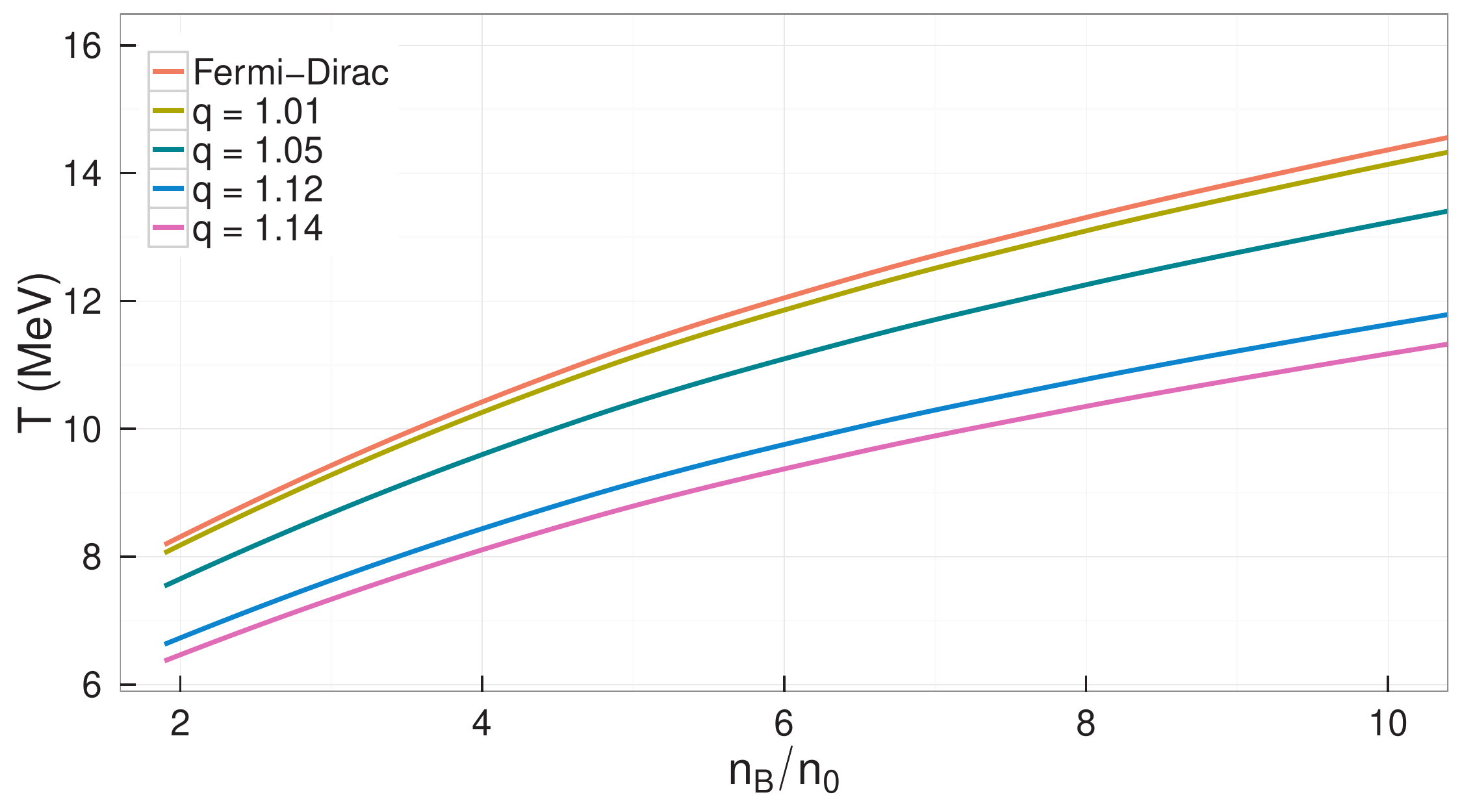}\label{fig3:a}}
\hfill
\subfloat[]{\includegraphics[width=0.7\linewidth]{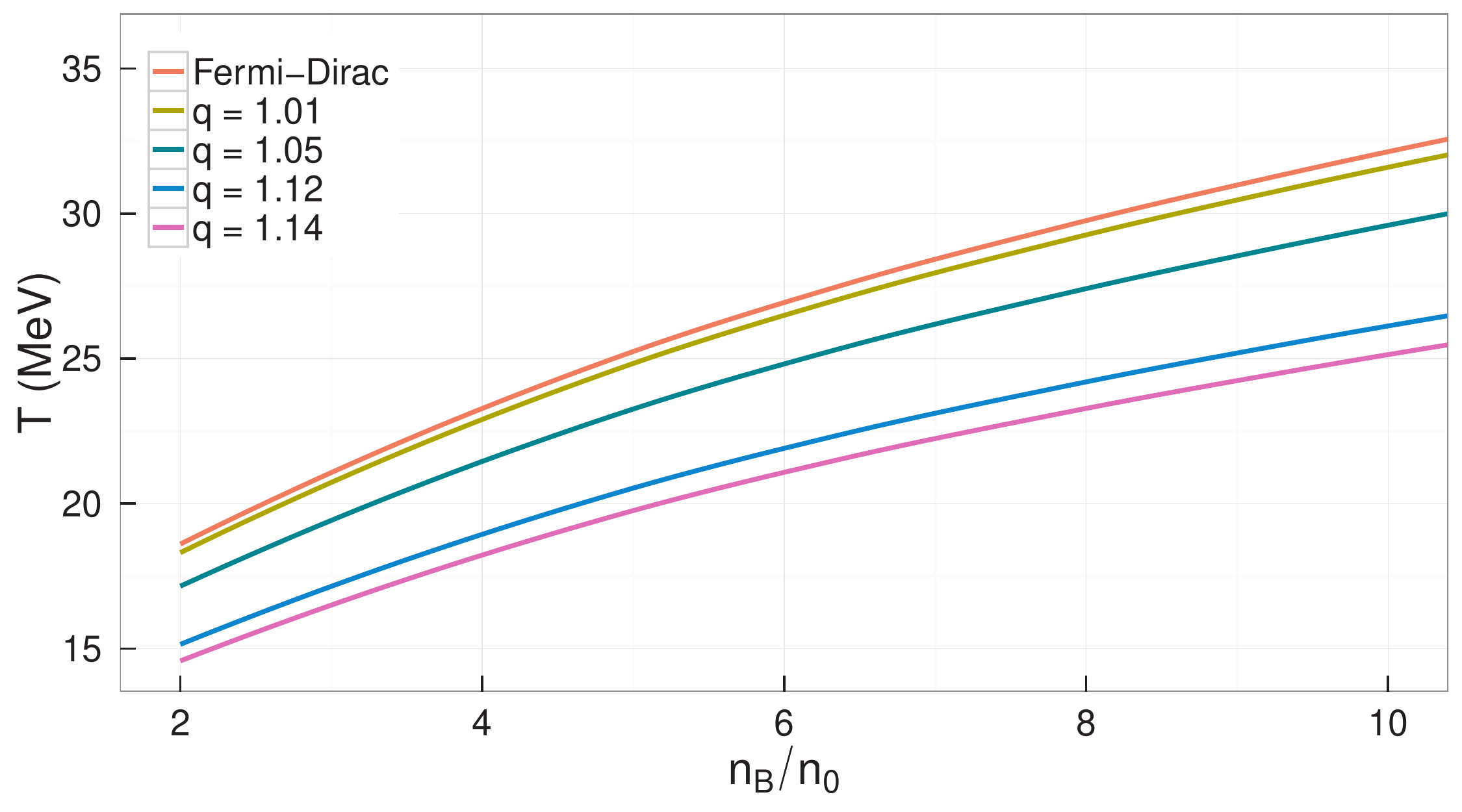}\label{fig3:b}}
\caption{Temperature as function of density (in units of nuclear
  matter saturation density) for different values of $q$ 
and a) ${\cal S}/n_B=1$, $Y_l=0.3$ and b) ${\cal S}/n_B=2$, $\mu_{\nu}=0$.} 
\label{fig3}
\end{figure*}

We then analyse the influence of non-additivity on the system temperature in Fig.\ref{fig3}. As $q$ increases, the resulting star internal temperature decreases. This behaviour was already found in \cite{Menezes:2014wqa}, but the temperatures involved in both snapshots are much lower here because the quark matter EOS is softer.

\begin{figure*}
\centering
{\includegraphics[width=0.7\linewidth]{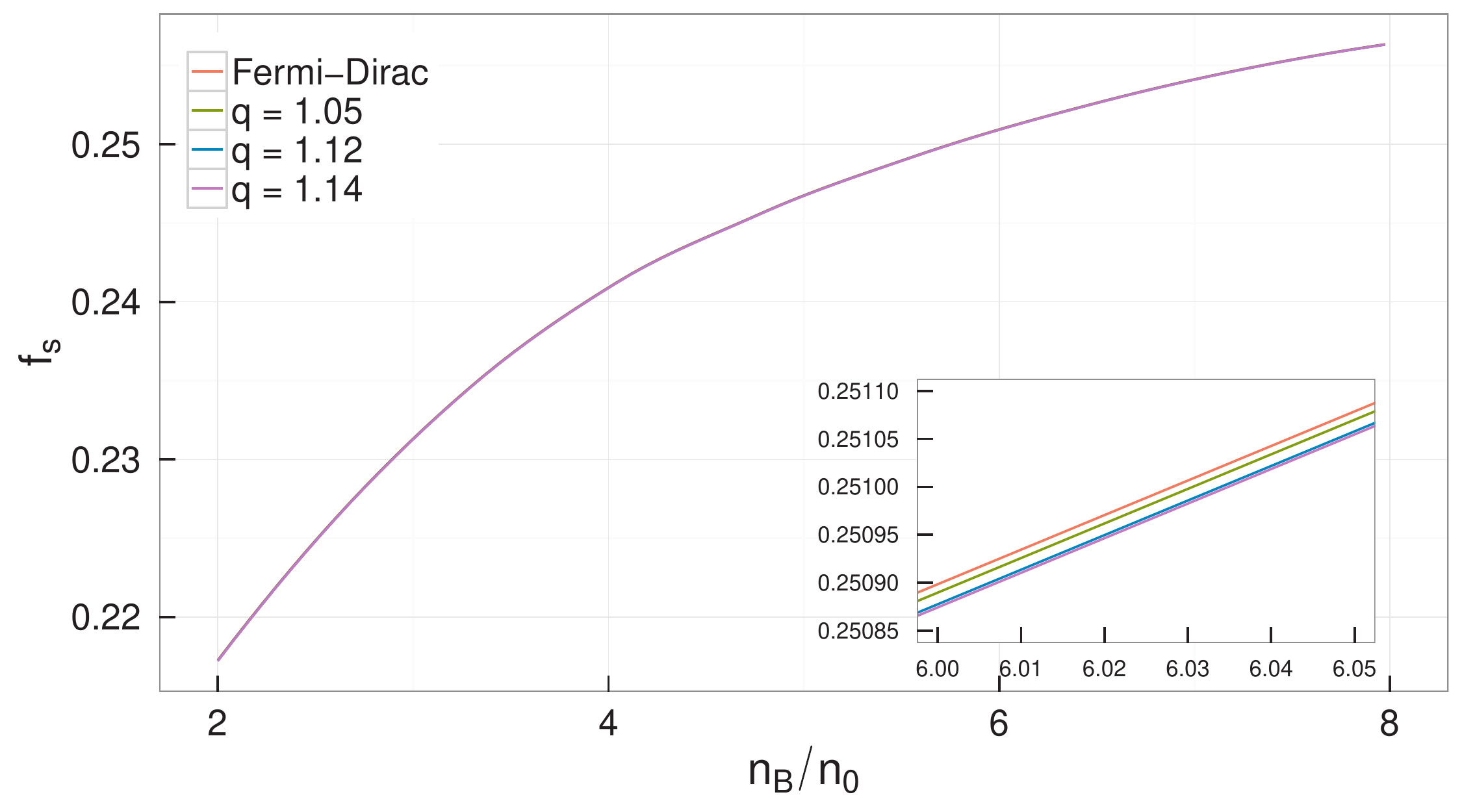}\label{fig4:a}}
\caption{Strange quark fraction obtained for ${\cal S}/n_B=1$,
  $Y_l=0.3$.}
\label{fig4}
\end{figure*}

Before we move to macroscopic properties, we observe that the amount of strangeness in the system is barely affected by the Tsallis statistics, as can be seem in Fig.\ref{fig4} for the first snapshot. If the second snapshot were plotted, one could see that the strangeness fraction would be slightly larger (reaches 0.33 at high densities), but the effect of the statistics remains negligible. This fact has consequences in the calculation of the star macroscopic properties, as discussed next.

\begin{figure*}
	\centering
{\includegraphics[width=0.7\linewidth]{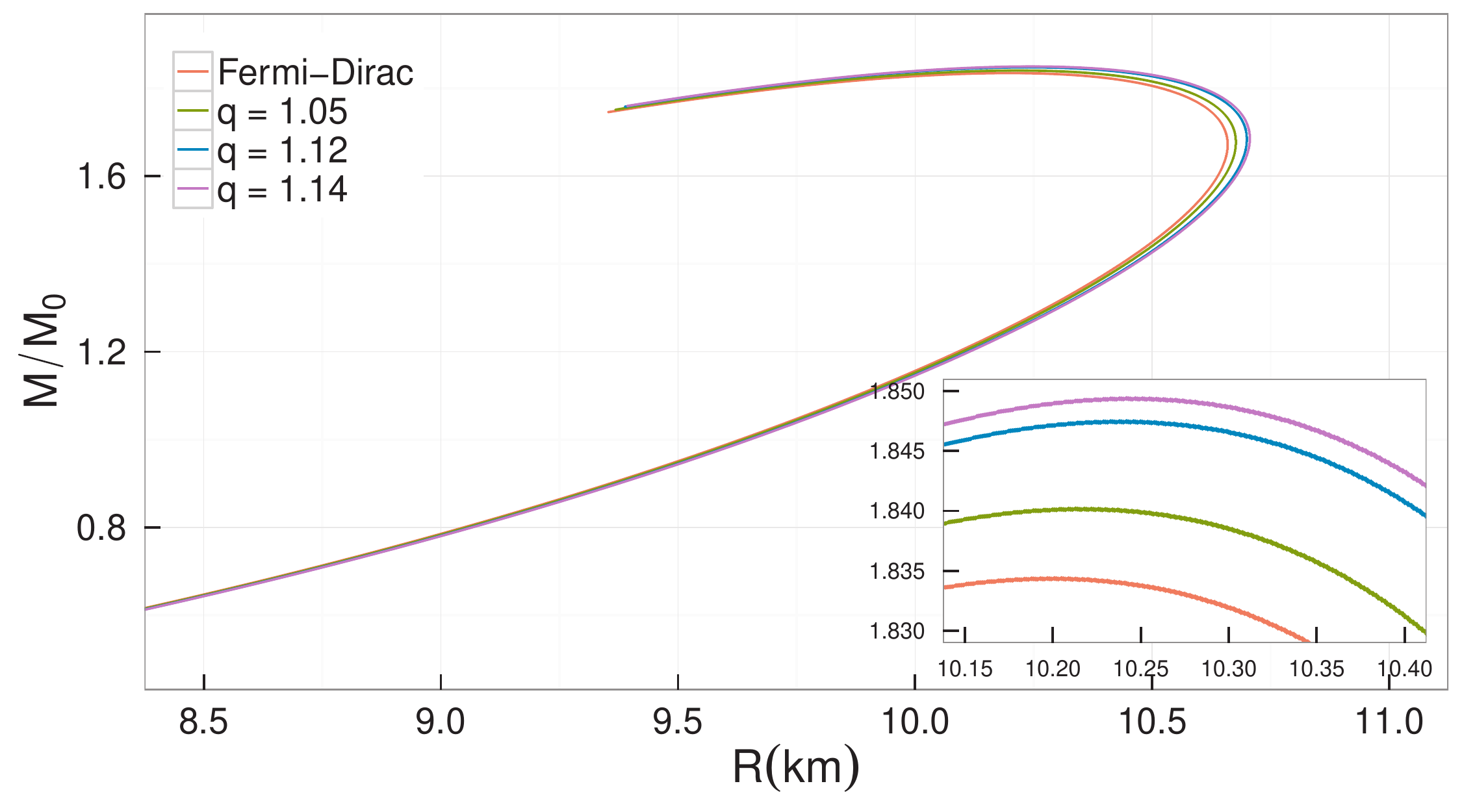}\label{fig:massRadius-b}}
	\caption{Mass-radius relation for 
          ${\cal S}/n_B=2$, $\mu_{\nu}=0$.}
\label{fig5}
\end{figure*}

Once the EOS is computed, it is used as input to the Tolman-Oppenheimer-Volkof equations \cite{Oppenheimer:1939ne}, which gives the macroscopic properties of interest that are displayed next in Fig. \ref{fig5} (just the second snapshot) and Table \ref{tab1}. In the same way as the results obtained in \cite{Menezes:2014wqa}, the final macroscopic properties are practically unchanged, but always tending to larger values for the masses and, in the present work, also for the radii. Therefore, the massive 2 $M_\odot$ stars  recently observed \cite{Demorest:2010bx,Antoniadis:2013pzd} cannot be described by simply increasing the value of the $q$-parameter. 
		
\begin{table*}[ht!]
\begin{center}
\begin{tabular}{|c |c |c |c |c|}
  \hline 
  Case & q & $M_{max}$/$M_{\odot}$& $Mb_{max}$/$M_{\odot}$ & $R(\rm{km})$\\
  \hline
  \multirow{4}{*}{ ${ S}/n_B=1$, $Y_{l}=0.3$ } & Fermi-Dirac &  1.8545 & 2.229  & 10.293\\ 
   & 1.05 &  1.8566 & 2.233 &  10.297  \\ 
   & 1.12 &  1.8591 & 2.238 & 10.302 \\ 
   & 1.14 &  1.8597 & 2.239 & 10.305\\   
  \hline
  \multirow{4}{*}{ ${ S}/n_B=2$, $\mu_{\nu}=0$ } & Fermi-Dirac &  1.8344 & 2.300 & 10.201 \\  
   & 1.05 &  1.8402 & 2.313 & 10.213 \\ 
   & 1.12 &  1.8475 & 2.329 &  10.233 \\ 
   & 1.14 &  1.8494 & 2.334&  10.242 \\   
  \hline
\end{tabular}
\end{center}
\caption{Stellar macroscopic properties: $M_{max}$ refers to the
  maximum gravitational mass, $Mb_{max}$ to its baryonic counterpart
  and $R$ to its radius.}
\label{tab1}
\end{table*}

\section{Final remarks}
The MIT bag model treats the strong interaction in a very naive way, i.e., through a constant. Hence, due to the simplicity of the model, we expected to see more clearly the effects of the Tsallis statistics. In fact, the modifications are always made at the level of a free gas. In contrast with the models used in \cite{Lavagno:2011zm, Menezes:2014wqa}, where the hadrons are mediated explicitly by mesonic fields, no information about the interaction between quarks and gluons is mimicked in the MIT bag model. 

The analyses of the first order transition in the QCD phase diagram shows, however, that the model is quite sensitive to non-extensivity. Unfortunately, the model cannot be used to calculate the CEP, as done in \cite{Rozynek:2015zca}, but the influence on the phase transition boundary is larger with the MIT than with the NJL model.

When the model is extended to incorporate stellar matter conditions and the macroscopic properties are computed, the effects are very small, but corroborate most of the findings obtained for neutron stars described by more sophisticated models in  \cite{Lavagno:2011zm, Menezes:2014wqa}, namely, maximum masses increase and internal stellar temperature decreases with the increase of the $q$ parameter.  In the present work, it is also clear that the radii also increase alongside the $q$ parameter, not an obvious result in the previous papers. 

\section{Authors contributions}
All the authors were involved in the preparation of the manuscript, read and approved the final version. This work was partially supported by CNPq under grants 300602/2009-0 (DPM), 310982/2014-6 (AD) and  435158/2016-3 (TNdS) and  PIBIC scholarship (PHGC). 

\bibliographystyle{spphys}       
\bibliography{refs-utf8.bib}   
\end{document}